\documentclass[12pt]{article}
\usepackage[protrusion=true,expansion=true]{microtype}
\usepackage{amsmath,amsfonts, amssymb, braket, bbold}
\usepackage{tikz}
\usetikzlibrary{trees,er,snakes,shapes,mindmap}
\usepackage{braket}
\usepackage{titlesec}
\titleformat{\section}
 {\normalfont\fontfamily{put}\fontsize{12pt}{14pt}\bfseries\color{black}}
{\thesection}{1em}{}
\titleformat{\subsection}
 {\normalfont\fontfamily{put}\fontsize{12pt}{14pt}\bfseries\color{black}}
{\thesection}{1em}{}
\def\sloppy{\tolerance=100000\hfuzz=\maxdimen\vfuzz=\maxdimen}
\def \beq  {\begin{equation}}
\def \eeq  {\end{equation}}
\def \beqar {\begin{eqnarray}}
\def \eeqar {\end{eqnarray}}
\allowdisplaybreaks
\def\sqr#1#2{{\vcenter{\vbox{\hrule height.#2pt
\hbox{\vrule width.#2pt height#1pt \kern#1pt
\vrule width.#2pt}\hrule height.#2pt}}}}

\def\S {{\cal S}}
\def\la {{\langle}}
\def\ra {{\rangle}}
\def\vx {{\vec x}}

\def\Tr {{\rm Tr}}

\def\vx {{\vec x}}

\def\del {\partial}

\def\bz {{\bar{z}}}

\def\vx {{\vec x}}

\def\del {\partial}

\def\bz {{\bar{z}}}

\def\A {{\cal A}}

\def\P {{\cal P}}
\def\T{{\cal T}}

\def\half{\textstyle{1\over 2}}

\begin{document}
\fontfamily{bch}\fontsize{11pt}{15pt}\selectfont
\def \CMP {{Commun. Math. Phys.}}
\def \PRL {{Phys. Rev. Lett.}}
\def \PL {{Phys. Lett.}}
\def \NPBProc {{Nucl. Phys. B (Proc. Suppl.)}}
\def \NP {{Nucl. Phys.}}
\def \RMP {{Rev. Mod. Phys.}}
\def \JGP {{J. Geom. Phys.}}
\def \CQG {{Class. Quant. Grav.}}
\def \MPL {{Mod. Phys. Lett.}}
\def \IJMP {{ Int. J. Mod. Phys.}}
\def \JHEP {{JHEP}}
\def \PR {{Phys. Rev.}}
\def \JMP {{J. Math. Phys.}}
\def \GRG{{Gen. Rel. Grav.}}
\begin{titlepage}
\null\vspace{-62pt} \pagestyle{empty}
\begin{center}
\vspace{1truein} {\large\bfseries
On Quantum Field Theory and Observers}\\
\vspace{6pt}
\vskip .1in
{\Large \bfseries  ~}\\
{\Large\bfseries ~}\\
{\sc V.P. Nair}\\
\vskip .2in
{\sl Physics Department,
City College of New York, CUNY\\
New York, NY 10031}\\
\vskip.1in
\begin{tabular}{r l}
{\sl E-mail}:&\!\!\!{\fontfamily{cmtt}\fontsize{11pt}{15pt}\selectfont vpnair@ccny.cuny.edu}\\
\end{tabular}

\vspace{.8in}
\centerline{\large\bf Abstract}
\end{center}
A generalization of the coadjoint orbit action describes the dynamics
of an observer (or instrument). We consider how this fits in with the view of observables in field theory being correlations of read-outs of instruments
and show how one recovers the usual $S$-matrix formulae.
A simple resolution of the Fermi paradox is also pointed out.

\end{titlepage}

\pagestyle{plain} \setcounter{page}{2}
\section{Introduction}

In recent works, Witten and collaborators brought in the observer 
in a nontrivial way in defining quantum field theory (QFT) on curved spacetimes \cite{{witten1},{witten2}}. In particular, Witten has argued that states corresponding to the observer can give a background-independent operator algebra which is important in the context of quantum gravity where the background geometry is subject to quantum fluctuations. For QFT on a flat or fixed background, i.e., without quantum gravity, the observer can be introduced but seems to play a fairly passive role. In fact, in quantum field theory as it is practiced,
for example, for the calculation of scattering amplitudes and decay rates,
we do introduce sources (or sinks) as a simple trick to collect correlators into a generating functional. Since the sources can be chosen or manipulated arbitrarily by agencies external to the system under study, they may be viewed as the mathematical expression of the observer in the practical version
 of QFT.
(This is not a particularly new observation. Local sources have long been
considered as the mathematical expression of instruments or detectors,
see, for example, \cite{Fred-Haag}.)
Many years ago, Schwinger tried to elevate sources as the primary objects of interest, fields arising only as an afterthought, as a convenient method, albeit immensely successful, of comparing or correlating sources \cite{schwinger1}.
No {\it a priori} reality is ascribed to fields. Unfortunately, Schwinger did not introduce any dynamics for the sources and, as a result, the ``source theory'' he constructed has generally been viewed as nothing more than a slightly different rewriting of standard QFT. However, in the light of the recent emergence of the key role of the observer in quantum gravity, we think it is worthwhile to reconsider this issue. We attempt a framework for augmenting QFT even on flat spacetime (or on a fixed background geometry) by introducing dynamics for sources and discuss the conclusions which might arise from it.\footnote{Schwinger also had an earlier theory of measurements in quantum mechanics where any instrument is viewed as a filter $M(a,b)$ which returns the value
``$a$" for an input ``$b$", with a certain probability \cite{schwinger2}. A chain of reasoning then leads to the identification
of $M(a,b)$ as $\braket{a|b}$, the inner product for the corresponding
states on a Hilbert space. This has more recently been developed into a groupoid formulation of quantum mechanics \cite{groupoid}. This is not the subject of our discussion, we are focused on quantum field theory.}

In the next section, we consider the probability that can be obtained or deduced from measurements which are recorded in terms of the settings of instruments.
This is followed by a discussion of the dynamics of the observer or instrument. 
We then comment on describing the instrument using density matrices,
and also on how the usual $S$-matrix formula fits into our formulation.
Finally we comment on how the Fermi problem, which has been 
analyzed by several authors in the context of signaling and Einstein causality in field theory, has a fairly simple resolution.

\section{What is observed}

Physics is about comparing and correlating the read-outs of different instruments. (Instrument, source/detector and observer are basically synonymous in what follows.)
Therefore we start by considering the dynamics of an instrument.
The instrument has a number of quantum states, say, $N$ of them.
(We will consider $N$ to be very large eventually.)
These states are characterized by a set of labels which can be viewed
as the eigenvalues of a compatible
set of observables. A compatible set in this context stands for 
a complete set of mutually commuting observables.
This is basically tautological since, if we start from the point of view of giving primacy to the instruments and accordingly we assign states 
with unique labels to it, the set of compatible
observables can be taken as
the set of all diagonal matrices in this basis of states.

We will assume that the read-outs corresponding to these 
different states can be distinguished in an unambiguous way. 
This may require the use of techniques to amplify certain signals.
For example, in the discussion in Doplicher \cite{doplicher},
there are three parts to the measurement.
There is the physical system being studied,
there is the microscopic apparatus which is treated quantum mechanically
and there has to be a macroscopic apparatus whose job is to ensure
that we can read out the initial and final states 
of the microscopic apparatus unambiguously.
The macroscopic apparatus is also taken to be quantum mechanical,
but with a large number of degrees of freedom, which provide sufficient
amplification of the measured signal to produce an unambiguous read-out.
The Heisenberg cut, in the usual terminology of measurements in
quantum mechanics, is somewhere in the transition from the
microscopic apparatus to the macroscopic one.
We consider these two together as constituting the instrument.

One of the advantages of considering just the instruments and their read-outs,
and physical theories as just a calculational framework
for understanding correlations among such read-outs, is that
one evades the concept of the collapse of the wave function. 
This can be illustrated by one of the often-used examples
involving the collapse of the wave function.
Consider the measurement of the spin of the electron along the $x_3$-direction followed by the measurement along the $x_1$-direction.
In the standard parlance, the wave function of the electron, in whatever state it was, collapses into an eigenstate of $S_3$, say, corresponding to
$S_3 = {\half}$ due to the first measurement.
The subsequent measurement then
collapses this state into an eigenstate of $S_1$, with probability
${\half}$ for each of the two choices, $S_1 = \pm {\half}$.
The terminology used here
 makes a presumption that there is reality to the electron
between the measurements and that it has some wave function
(say corresponding to $S_3 = {\half}$) into which it ``collapsed" after the first measurement, and in which state it continues to be
thereafter until the second measurement.
None of this is directly observed.
Therefore, stripping this down to the bare observations, what we really have is that, at a certain time,
instrument 1 is in a certain state 
(which may be interpreted as corresponding to
$S_3 = {\half}$) and this evolves to
instrument 2 showing read-outs at a later time, the two possibilities for
instrument 2 (which may be interpreted as corresponding to
$S_1 = \pm {\half}$) being of equal probability.
There is a probability of ${\half}$ that the initially
obtained reading of instrument 1 leads to the read-out corresponding to
$S_1 = {\half}$ for instrument 2, and there is a probability 
of ${\half}$ that it will lead to a read-out
corresponding to $S_1 = - {\half}$ for instrument 2.
The correlation of read-outs of instruments 
over time is probabilistic, but
the read-out for any measurement,
 even a single one, is definite.
No state of the electron is presumed in this observation and so
the concept of
``collapse" is not directly relevant.\footnote{One might argue that the
``collapse" does occur for the microscopic apparatus, and it is
what leads to
definite read-outs. For this reason, we have used the word ``evade" 
regarding collapse for the system under study, it may not be eliminated
but transferred to the apparatus. Anyway, this is not the main point for us.
We are interested in an action for the instrument and how it fits into what we can observe.}
The role of the physical theory is to understand the laws governing these 
correlations among instruments.
Our aim is to express the standard calculational framework of quantum field theory in terms of correlations of 
read-outs of instruments
and their dynamics.
But first, we will use the conventional formalism of field theory to
define the quantity of interest.

A typical observation would run as follows.
We start with some state $\ket{\alpha}$ as the state of the world, say
at time $t = t_0$.
At some point in time, say $t_1> t_0$, an instrument is turned on
with a certain setting $\{ a\}$ for the read-outs.
Consider then the further time-evolution of this set-up
to $t > t_1$. The state at time $t$ is of the form
\beq
\ket{\alpha, a , t} = U(t, t_1) \, f_a (t_1) \, U(t_1, t_0) \ket{\alpha}
\label{obs01}
\eeq
where the insertion of $f_a (t_1)$ characterizes the turning-on of
the instrument at time $t = t_1$ and $U$ is the time-evolution operator.
(If we think of the instrument as made of a collection of particles which can be described by quantum fields, we may think of $f_a$ as a composite field operator creating the instrument in a state labeled by $\{ a\}$.)
At this time, the instrument can show a read-out $\{b \}$
or another set of definite values $\{b'\}$,  etc.
We now ask the question: What is the probability that 
the set-up in (\ref{obs01}) 
is identical to instrument showing read-out
$\{ b \}$ at some time $t_2 < t$?
The instrument showing read-out $\{ b\}$ at $t_2$ is described by
\beq
\ket{\beta , b, t} = U(t, t_2) f_b (t_2) \, U(t_2, t_0)\ket{\beta}
\label{obs02}
\eeq
Notice that the final state of the world
in (\ref{obs01}) could have changed in the process of measurement,
so we should match with read-out $\{b\}$ for the instrument,
which is definite, but with any state $\ket{\beta}$ for the world.
This is why we allow $\beta$ to be different from $\alpha$, in general.
The probability amplitude for obtaining (\ref{obs02}) from
state (\ref{obs01}) is
\beq
\A_{\beta b, \alpha a} =
\bra{\beta} U^\dagger (t_2, t_0)  f_b^* (t_2) U^\dagger (t, t_2)
U(t, t_1) f_a (t_1) U(t_1, t_0) \ket{\alpha}
\label{obs03}
\eeq

If we view $f_a(t_1)$ as the operator creating the instrument at
time $t_1$, then the amplitude (\ref{obs03}) is the propagator for the instrument
on a time-contour from $t_0$ to $t$ and then back to $t_0$, with
$f_b^*$ on the reverse branch of the contour. (The state of the world changes as well.)
Notice that if $t_2 > t_1$, then we can use $U^\dagger (t, t_2) U(t, t_1)
= U(t_2, t_1)$ and write
\beq
\A_{\beta b, \alpha a} =
\bra{\beta} U^\dagger (t_2, t_0)  f_b^* (t_2) 
U(t_2, t_1) f_a (t_1) U(t_1, t_0) \ket{\alpha}
\label{obs03a}
\eeq
The propagator for the instrument is then on the first branch. This 
simplification is not possible if we have several instruments 
making measurements at different times.
For example, for two instruments, the amplitude is of the form
\beqar
\A_{\beta b_1 b_2, \alpha a_1 a_2} &=&
\bra{\beta} U^\dagger (t_4, t_0)  f^*_{b_1} (t_4) 
U^\dagger(t_3, t_4) f^*_{b_2}(t_3) U^\dagger (t, t_3)\nonumber\\
&&\hskip .2in\times U(t, t_2) f_{a_2} (t_2) U(t_2, t_1) f_{a_1}(t_1) U(t_1, t_0) \ket{\alpha}
\label{obs03b}
\eeqar
If $t_3 > t_2$, we may use $U^\dagger (t, t_3) U(t, t_2)
= U(t_3, t_2)$, but the amplitude will still have propagators involving
$f^*_{b_1}(t_4)$ on the reverse branch.

The time-evolution in (\ref{obs01}), (\ref{obs02}) will
involve the time-evolution of the instrument as well as
the rest of the world.
(``The rest of the world" would include any fields which interact with
the instruments.)
In general, there will be entanglement between 
the state of the instrument and the rest of the world as well.
Further, in general, the state of the world must be described by a density
matrix.
Denoting the probability for $\ket{\alpha}$ as $p_\alpha$, 
the probability we are interested in is given by
\beq
\P (b, a) = \sum_\beta \sum_{\alpha} \vert \A_{\beta b, \alpha a}\vert^2
p_\alpha
\label{obs04}
\eeq
This is the quantity of interest. While the instruments can show
read-outs labeled as $a$ and $b$, the rest of the world can evolve into any state, which is why we sum over $\beta$ as well.

Each propagator for the instruments in $\P( b, a)$ can be expressed as 
a path-integral over the world line of the instrument
coupled to any fields which may exist in the ambient spacetime.
Our aim is therefore to introduce the action for the instrument and rewrite
$\P (b, a)$ as a functional integral. This will give
a prescription for correlations of read-outs without directly using
states on a Hilbert space.
\section{Dynamics of the observer}
Turning to the question of how $\P (b,a)$ can be calculated, notice that
in the language of quantum field theory, since
the instrument is made of a number of particles which can be 
created from the vacuum by the action of creation operators, we can
introduce the standard annihilation/creation operators $c_i$ and
$c^\dagger_i$, which are presumably composites of more fundamental particle operators, with $c^\dagger_i$ creating the instrument in the state
$\ket{i}$ as $\ket{i} = c^\dagger_i \ket{0}$. 
However, we will simply postulate the existence of the instrument with
its internal states, in the spirit of giving primacy to the observer.

The state $\ket{i}$ can presumably evolve
to other states of the instrument, so the Hamiltonian is of the form
$H = c^\dagger_k \,H_{kl} \, c_l$, where
$H_{kl}$, $k, l = 1, 2, \cdots, N$, is an $N\times N$ matrix.
It is useful to consider the time-evolution in terms of coherent states
for $c_i$, $c^\dagger_i$, with $f_i (z) = \bra{z} c_i^\dagger \ket{0}
= \bz_i \, e^{ - {\half} \bz\cdot z}$.
The path-integral for the time-evolution of the instrument then takes the form
\beqar
\bra{j} e^{-i H t} \ket{i} &=&
\int {d^2 z \over \pi} {d^2 z' \over \pi}\, f^*_j (z) \int  [Dz D\bz] 
\, e^{i \S (z, z', t)} \,  f_i (z')\nonumber\\
\S (z, z', t) &=&\int_0^t dt'  \left[{i \over 2} \bigl({\dot \bz}_k z_k - \bz_k {\dot z}_k 
\bigr)
- \bz_k H_{kl} z_l\right]
\label{obs1}
\eeqar
(Here we are considering the standard time-evolution of a pure state
generated by a Hamiltonian. In general, the instrument will be in
a mixed state. We will consider some features of such a situation decsribed by a density matrix in section 4.)
In (\ref{obs1}), we have the functional integration over
all $z$, $\bz$ along the trajectory connecting $z', \bz'$ at $t = 0$
to $z, \bz$ at time $t$. This is indicated by the functional measure
$[Dz D\bz] $. The integration over the initial point $z', \bz'$ and the
final point $z, \bz$ have to include the choice of states $\ket{i}$,
$\ket{j}$, with the coherent state wave functions
$f_i (z')$, $f^*_j (z)$. This is separately indicated in (\ref{obs1}).

With the given Hamiltonian, $c^\dagger_k c_k$ is conserved, and, since we are interested in the evolution of just one copy of the instrument, we can set
$c^\dagger_k c_k =1$ as a constraint. 
This is equivalent to $\bz_k z_k - 1 = 0$.
The conjugate constraint
fixes an overall phase among the $z_k$'s, so that
(\ref{obs1}) reduces to a path-integral
on $\mathbb{CP}^{N-1} = SU(N)/U(N-1)$.
Without loss of generality, we can parametrize this as
$z_k = U_{k 1}$, $U_{k l} \in SU(N)$.
The result is
\beq
\S = \int dt\, \left[ -i (U^\dagger {\dot U})_{11} - (U^\dagger H U )_{11}
\right] = \int dt\, \Tr \left[ h\,U^\dagger \left( - i {\del \over \del t} - H
\right) U \right]
\label{obs2}
\eeq
where $h = \ket{1} \bra{1}$. The functional integration measure reduces to
the standard volume element on $\mathbb{CP}^{N-1}$
at each point in time in defining (\ref{obs1}) by time-slicing.

The instrument is at some location in space, idealized as the point
with spatial coordinates $\vx$. The result
(\ref{obs1}), (\ref{obs2}), where we consider only transitions among the internal states of the instrument, applies to the rest frame of the instrument.
In seeking the generalization of this to the fully relativistic case, there are two
points to keep in mind. The states are not necessarily degenerate so the square of the mass is an $N\times N$ positive matrix.
(There may be degeneracies among subsets of states.)
This matrix can be taken to be diagonal. To motivate the possibility of
including scalar fields in the ambient space which can interact with the instrument, it is useful momentarily to think of a field $\chi_i$ as representing the instrument with an action of the form
\beq
\S = \int \left[ {1\over 2} (\del \chi)^2 - {1\over 2} \chi (m^2) \chi 
- {1\over 2} \chi\,\Phi\, \chi \right]
\label{obs2a}
\eeq
where $\Phi$ is a scalar field (or composites thereof).
Thus $m^2 \rightarrow m^2 + \lambda \Phi$ can be used to include this type of interaction.
$m^2 + \lambda \Phi$ is also an $N\times N$ matrix, not diagonal in general.
We will need the square-root of this expression for use in the
action. In the coherent state basis, this leads to
$\Tr \left[h U^\dagger \Bigl( \sqrt{m^2 + \lambda \Phi } \Bigr)U\right]$.
Assuming there is a nonzero diagonal part $m_0^2$
(which will be the case of interest to us),
we can write $m^2 = m_0^2 + \delta m^2$ and expand
\beqar
\Tr \left[h U^\dagger \Bigl( \sqrt{m^2 + \lambda \Phi } \Bigr)U\right]
&\equiv& \Tr \left[ h U^\dagger \, M(z, \bz, x) U \right]\nonumber\\
&\approx& \Tr \left[h U^\dagger \Bigl( m_0 + {1\over 2} m_0^{-1} (\delta m^2 + \lambda \Phi)  + \cdots\Bigr)U\right]
\label{obs2c}
\eeqar
The second point is about
the need to maintain the invariance under reparametrization of
$\tau \rightarrow \tau' = \tau'(\tau)$ for the world line $x^\mu (\tau)$
of the instrument. It is now easy to see that the relativistic generalization
of (\ref{obs2}) is given by
\beqar
\S_{\rm instr} &=& \int d\tau \, \Tr  \left[h\,U^\dagger\left( - i  {\del U \over \del \tau}
-  A_\mu {d x^\mu \over d \tau } \, U - M \, {d s \over d\tau} \,U\right) \right]
\label{obs3}\\
&=& \int d\tau \, \Tr  \left[h\,U^\dagger\left( - i  {\del U \over \del \tau}
-  A_\mu {d x^\mu \over d \tau } \, U\right) \right]
+ \S_0
\label{obs3a}
\eeqar
where, in (\ref{obs3a}),  we note that the $M$-dependent term
in (\ref{obs3}) can be rewritten using
\beq
\S_0 = 
= -\int d\tau \left[{{\dot x}^2\over 4 e} + \bigl(\Tr(h U^\dagger M U)\bigr)^2 e
\right]
\label{obs3b}
\eeq
Here $e$ is a one-dimensional frame field; eliminating it leads to the
usual form of the action given in (\ref{obs3}), namely,
\beq
-\int \Tr (h U^\dagger M U) \sqrt{{\dot x}^2} d\tau = -\int \Tr (h U^\dagger M U) \, ds
\label{obs3bb}
\eeq
If we use the gauge fixing $x^0 = \tau$ for the reparametrization symmetry
and neglect ${\dot x}^i$, as would be appropriate for large masses
(or the rest frame of the instrument), (\ref{obs3}) reduces to (\ref{obs2})
as required. In this case $H$ is a combination of $A_0$ and
$\sqrt{m^2 +\lambda \Phi}$.
(The term $\S_0$, for a single mass with no interaction term
$\lambda \Phi$ is what was used in \cite{witten2}.)
In the following, we regard $A_\mu$ and $\Phi$ as fields in the ambient spacetime.

The action in (\ref{obs3})  is a generalization of
the co-adjoint orbit action used for a particle with
$U(N)$ charges coupled to a gauge field $A_\mu$ which
leads to the Wong equations of motion\cite{wong}.
If we neglect $\Phi$ and take all masses to be the same, it is exactly
the action for the Wong particle; the generalization involves allowing for
$M$ to be a general matrix, just as the $A_\mu$ are.
In the general case, with 
nondegenerate masses and the coupling to
$\Phi$, the action (\ref{obs3}) does not
have the full gauge symmetry $(-iA)_\mu 
\rightarrow  g (-i A_\mu) g^{-1} - \del_\mu g g^{-1}$,
$U \rightarrow g U$; it still has symmetry for a subgroup
of products of $U(n)$'s corresponding to the
invariance $g^\dagger (m^2 + \lambda \Phi ) g = (m^2 + \lambda \Phi )$.
(We may still consider $A_\mu$ as a $U(N)$ gauge field,
but spontaneously broken in the ambient spacetime to
the invariance subgroup of $m^2 +\lambda \Phi$.)
Notice that the first term in 
(\ref{obs3a}) is a topological term
independent of the background metric.
We could view $M ds $ as part of the connection one-form,
but it is metric-dependent via $ds$, namely,
with a nontrivial metric tensor $g_{\mu\nu}$ for the ambient spacetime, 
$ds = \sqrt{g_{\mu\nu} dx^\mu dx^\nu}$.
Also gauge transformations do not change $M$, in this sense it is not exactly
a connection one-form. The action
(\ref{obs3}) also shows that $M$ is conjugate to the proper time
$s$ if we consider a canonical quantization.
Our claim is that this generalization of the Wong action \cite{wong}
is what is needed for an observer/instrument 
with internal degrees of freedom.
The short-distance operator product expansion for the operators on the world line of the observer is the background-independent algebra considered in
\cite{witten2}. In the present context, these will be
operator products like $A_\mu (x(\tau)) A_\nu (x(\tau'))$,
$\Phi (x(\tau)) \Phi (x(\tau'))$, etc.

We can now consider the functional integral representation
of $\A_{\beta b, \alpha a}$
in (\ref{obs03}). We will need functional integrations
over  $A_\mu$ and $\Phi$, as these are fields in the ambient spacetime.
But first, it is useful to recall that the path-integral
representation of the usual Wilson line has the form \cite{wilsonline}
\beqar
W_{ba} (L) &=& \left[P \, \exp\left( - \int_L A_\mu dx^\mu\right)\right]_{ba}
\nonumber\\
&=& \int [Dz D\bz]\, f_b^* (z,\bz)\, e^{i \S_L} \, f_a (z',\bz')
\label{obs3c}\\
S_L &=& \int_L \Tr h U^\dagger \left( - i {\del U \over \del \tau}
- A_\mu {dx^\mu \over d\tau} U \right)
\nonumber
\eeqar
Here $L$ denotes the path along which the integration is done
and $P$ denotes path-ordering of $A_\mu$ as usual.
For our case, we also have $M ds$ as part of the connection, as
in (\ref{obs3}), so, for the instrument we define the path-ordered exponential
integral
\beqar
I_{ba} (L) &=& \left[P \, \exp\left( - \int_L (A_\mu dx^\mu + M ds) \right)\right]_{bx,ax'}\nonumber\\
&=& \int [Dz D\bz] f_b^* (z, \bz) \, e^{i \S_{\rm instr}(L)}\, f_a (z',\bz')
\label{obs3d}\\
\S_{\rm instr}(L) &=&
\int_{x', z', \bz'}^{x, z, \bz}  d\tau \, \Tr  \left[h\,U^\dagger\left( - i  {\del U \over \del \tau}
-  A_\mu {d x^\mu \over d \tau } \, U - M \, {d s \over d\tau} \,U\right) \right]
\nonumber
\eeqar
While $z, \bz$ and $z', \bz'$ are integrated over with
$f_b^*$ and $f_a$, we consider fixed values of the final and initial
positions, $x$ and $x'$, of the instrument.

The amplitude in (\ref{obs03}) involves the evolution
of the instrument starting at time $t_1$ to $t$ and then evolution backward
to $t_2$ which is the time of the second read-out.
Thus we will need to consider paths $L$ in (\ref{obs3d}) 
as starting at $x^0 = t_1$, going forward to $t$ and then back to $t_2$.
When this is used in (\ref{obs03}), the time-coordinate
for the action integral for the fields $A$, $\Phi$ will  
go from $t_0$ to $t$ and then back to $t_0$, but the line integral
in $I_{ba}$ is only over a segment of this contour, from $t_1$ to $t$ to $t_2$.
We can take $(t-t_2)$ to be large or as small as we like.
Using a functional integration for the fields as well, we can now write the amplitude as
\beqar
\A_{\beta b, \alpha a} &=&\int [DA D\Phi Dx] \braket{\beta| A, \Phi} 
e^{ i \S_{\rm C} (A, \Phi) }\, I_{ba}(L) \, \braket{A', \Phi'|\alpha}
\nonumber\\
&=& \int [DA D\Phi Dx D\mu] \braket{\beta| A, \Phi} \,
e^{i\S_{\rm C} (A, \Phi)} f_b^* \,e^{i \S_{\rm instr}} \, f_a \, 
\braket{A', \Phi'|\alpha}
\label{obs3e}
\eeqar
Here $\S_{\rm C} (A, \Phi)$ is the action for the fields $A$, $\Phi$
in the ambient space.
It connects $A',\Phi'$ to
$A,\Phi $ over the range of the time-integration.
It is defined on a time-contour from $t_0$ to $t$ and back to $t_0$ as mentioned above.
Although this is similar to the Schwinger-Keldysh (SK) contour, there is
a distinction. Here the contour arises just for the amplitude
whereas the usual context for the SK contour is for time-evolution
of the density matrix.
The functional integration over $A$ should be defined using the product of the volume 
element of the gauge-orbit space at each point in time, 
in a time-sliced version of the functional integral, as is usually done.
Further, $[D\mu]$ denotes integration over $z$, $\bz$ with the volume element
of $\mathbb{CP}^{N-1}$ at each point in time.

\begin{figure}[!t]
\begin{center}
\scalebox{1.2}{\includegraphics{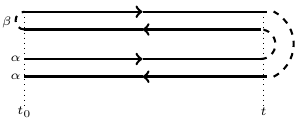}}
\caption{The SK$_4$ contour. The inner two branches are for $\A$, the outer two branches (3 and 4) are for $\A^*$. The time-intervals are from $t_0$ to
$t$; the contours are extended slightly with arcs attached
for clarity of the picture.}
\label{fig1}
\end{center}
\end{figure}

In the expression for the probability $\P (b, a)$, we have 
$\A^*_{\beta b, \alpha a}$, which we can again write as a path-integral
of the form (\ref{obs3e}), but with reversal of the time-contour
(as needed for the conjugate amplitude).
Thus the full path-integral for $\P(b, a)$ will be over a Schwinger-Keldysh-like
contour, with four branches rather than two;
we denote this contour by SK$_4$, see Fig.\,\ref{fig1}.
The inner two branches in the figure, say branches 1 and 2, correspond to
$\A_{\beta b, \alpha a}$ while the outer ones, say branches 3 and 4,
correspond to the
conjugate $\A^*_{\beta b, \alpha a}$.
We also have summation over $\beta$.
This means that the fields $A$, $\Phi$ can be continued 
on to the third and fourth branches 
and the action $\S (A, \Phi )$ is integrated over the full contour
SK$_4$. The expression for $\P(b, a)$ is thus
\beqar
\P (b, a) &=& \int [DA D \Phi ] [ Dx \, D{\tilde x}]\,e^{i \S_{\rm SK_4}(A, \Phi)} 
 I^*_{ba} ({\tilde L}) \, I_{ba} (L) \,  p(A', \Phi'; {\tilde A}', {\tilde \Phi}')
  \nonumber\\
&=&\Big\langle \int [Dx\, D{\tilde x}] \, I^*_{ba}({\tilde L})\, I_{ba} (L)
\Big\rangle
\label{obs4a}
\eeqar
For the third and fourth branches of the SK$_4$ contour, we designate
the variables with a tilde. These are independent of
the ones on the first two branches.
The angular brackets in the second line
of (\ref{obs4a}) indicate the integration over all the fields 
$A$, $\Phi$, ${\tilde A}$, ${\tilde \Phi}$, etc.
with $e^{i \S_{\rm SK_4}(A, \Phi)}\, p(A', \Phi'; {\tilde A}', {\tilde \Phi}')$.
Finally, we have
\beq
 p(A', \Phi'; {\tilde A}', {\tilde \Phi}')
= \sum_\alpha \braket{A', \Phi'|\alpha} 
p_\alpha 
\braket{\alpha| {\tilde A}', {\tilde \Phi}'}
\label{obs4b}
\eeq
This gives the density matrix for
the initial state of the world in terms of
the values $A'$, $\Phi'$ at the begining
and ${\tilde A}' $, ${\tilde \Phi}'$ at the end of the
SK$_4$ contour.

When we consider several observers or several instruments, say $M$ of them,
we get
\beqar
\P (\{b\}, \{a\}) &=&
= \Big\langle\int \prod^M_i [Dx_i\, D{\tilde x}_i ]\,
 I^*_{b_1 a_1}({\tilde L}_1) \cdots I^*_{b_M a_M}({\tilde L}_M)
\nonumber\\
 &&\hskip .7in
 I_{b_1 a_1}(L_1) \cdots I_{b_M a_M}(L_M)
\Big\rangle
\label{obs5}
\eeqar
The different instruments may have different numbers of 
internal states, say, $N_1$, $N_2$, $\cdots$, $N_M$. If
$N_{\rm max}$ is the highest of these values, the field
$A_\mu$, $\Phi$ will be in some subalgebra of
$U(N_{\rm max})$, the choice of this subalgebra being a 
defining characteristic
of the instrument.

Equation (\ref{obs5}) expresses the basic observable $\P (\{b\}, \{ a\})$
as a correlation among the read-out histories 
or among a set of generalized Wilson lines
of the instruments.
No reference to states on a Hilbert space is needed at this stage.
The correlators themselves are calculated using the action for the fields $A_\mu$,
$\Phi$ in the ambient spacetime.
But ultimately, as mentioned in the Introduction, we can view
these fields merely as a calculational tool
for the correlations of instrumental read-outs, with no direct observable 
reality attributed to them. In this sense, (\ref{obs5})
is the expression of Schwinger's goal in his ``source theory".
Equation (\ref{obs5}) is the main result of this paper.

This equation expresses observables in terms of the world lines
of the instruments and in this sense it has some conceptual similarity to a Dirac-Feynman histories-based approach to quantum mechanics. Such an approach has also been argued to be 
the appropriate framework to include quantum gravity
\cite{har-sor}.
Also equation (\ref{obs5}) has some similarity to the source-detector
approach used in \cite{dickinson}.
There, the authors consider the expectation values of
local operators. The ``sensitivity" of the detector to such a
measurement is then shown to involve only retarded commutators,
in agreement with Einstein causality.
Our formula is different in making the instrument more central
and also involves the SK$_4$ contour rather than the usual
SK contour.
However, if we consider all insertions of $f$'s and $f^*$'s to be
on the first branch only, and hence the conjugate insertions
are on the fourth branch, the second and 
third branches can be removed
using the composition property of time-evolution and
the contour SK$_4$ will reduce to the usual SK contour.

In the rest of this paper, we make some comments
related to (\ref{obs5}).
In general, the initial labels $\{ a \}$, rather than being definite, will follow
a probability distribution. Secondly, the standard $S$-matrix
analyses can be obtained by suitable approximations to
(\ref{obs5}). We will comment on these in turn. We will also briefly indicate how Einstein causality is obtained in using
(\ref{obs5}) for the much-discussed Fermi problem/paradox.

\section{Density matrix for the observer}
If we consider the instrument as a quantum mechanical system
placed in an environment
with various fields, it is clear that the state of the instrument has to be described by a mixed state.
Even if we assume the initial settings can be controlled
and distinguished from one another, there is a nonzero probability,
perhaps very small,
for settings other than the intended one.
This can be due to interactions with fields in the environment
or due to entanglement between the instrument and ambient fields,
which can give a mixed state reduced density matrix for the instrument.
Further, if we consider the background 
spacetimes with a horizon and a temperature, the instruments will
reach thermal equilibrium and hence they should again be 
described by a density matrix.
Even though we refer to fields which may exist
in the ambient spacetime, these fields,
and even the nature of the spacetimes in which the instruments reside,
are to be
considered as an interpretation inferred from
the correlations of records made by the instruments;
this is the point of view we have tried to emphasize. 
Nevertheless, to take account of such possibilities,
one should describe the initial state of the instruments 
by a density matrix, a mixed state in general.

The reduced density matrix, because of couplings to the ambient spacetime,
will obey a version of the GKSL equation \cite{gksl}. Just to be specific, 
for example,
for a heavy instrument coupled to
scalar field maintained at a temperature $T$ in the ambient spacetime
(for which we can use the approximation in the second line of (\ref{obs2c})\,), the GKSL equation for the reduced density
matrix is given by
\begin{align}
{\del \rho_{ab} \over \del t} =
  2 \pi \lambda^2 \sum {1\over 2 \omega_k V}
\Bigl[& -T^\alpha_{ac} T^\alpha_{cd}\, \rho_{db}~ (n_k+1) \delta (E_c -E_d +\omega_k) \nonumber\\
& -T^\alpha_{ac} T^\alpha_{cd}\, \rho_{db} ~n_k \delta (E_c -E_d -\omega_k) \nonumber\\
& - \rho_{ad}\,T^\alpha_{dc} T^\alpha_{cb}~ ( n_k+1) \delta (E_c -E_d+\omega_k) \nonumber\\
& - \rho_{ad}\, T^\alpha_{dc} T^\alpha_{cb}~ n_k \delta (E_c -E_d -\omega_k) \nonumber\\
& + T^\alpha_{ac}\, \rho_{cd}\,T^\alpha_{db}~ n_k \delta (E_d -E_b +\omega_k) \nonumber\\
& + T^\alpha_{ac}\, \rho_{cd}\,T^\alpha_{db}~ (n_k+1) \delta (E_d -E_b -\omega_k) \nonumber\\
& + T^\alpha_{ac}\, \rho_{cd}\,T^\alpha_{db}~ n_k \delta (E_a -E_c -\omega_k) \nonumber\\
& + T^\alpha_{ac}\, \rho_{cd}\, T^\alpha_{db} ~(n_k +1) \delta (E_a -E_c +\omega_k) \Bigr]
\label{E31}
\end{align}
Here $V$ is the spatial volume and we can take 
the large volume limit in the usual way.
The scalar field coupling is of the form
$\Phi = 2 m_0 \, \phi^\alpha T^\alpha_{ab}$, it
 carries out the transitions
$a \longleftrightarrow b$ upon emission or absorption of the
$\phi$-particles. $T^\alpha_{ab}$ are matrices effecting the transitions
and can be viewed as the jump operators in the GKSL language.
Also $\lambda$ is the strength of coupling between $\phi^\alpha$
and the instrument.
We have taken the fields $\phi^\alpha$ to obey thermal correlations, with
\beq
n_k = {1\over e^{\omega_k/T} - 1}
\label{E31a}
\eeq
where $T$ is the temperature. (We have not gone over the derivation
of (\ref{E31}), since, apart from the fact that we use 
thermal correlations for the fields $\phi^\alpha$, the 
logical steps in obtaining it follow the
standard derivation of the GKSL equation.)
It is easy to see that
the choice
\beq
\rho_{ab} = {1\over Z} e^{- E_b/T} \delta_{ab}
= \rho_b \, \delta_{ab}
\label{E32}
\eeq
leads to vanishing of all terms on the right hand side of (\ref{E31})
leading to the equilibrium solution.
This clearly shows the thermalization of the instrument
and the need for using a density matrix for it
unless the energy differences for the different readings of the instrument
are large compared to $T$.
Even if we do not consider fields at nonzero temperature,
the state of the instrument will be described, in general, by a 
mixed state due to entanglement and interactions with the 
environment.

The state of the instruments which is obtained upon completion of a set of measurements is more subtle.
The results of the measurements is $\P(\{b\}, \{a\})$, where the labels
$\{ b\}$ correspond to a set of compatible observables
(or a set of mutually commuting set of operators).
In fact, $\P(\{b\}, \{a\})$ is inferred from measured values
$\la A_i\ra$ which are taken to be 
$\la A_i \ra = \sum_{b,a} (A_i)_b \, \P(\{b\}, \{a\})\rho_a$ for a set of mutually commuting 
observables $A_i$. By assigning a set of states $\{ \ket{b}\}$ to the instrument, this may be expressed as $\la A_i \ra = \Tr (A_i \rho' )$,
$\rho' = \sum_{b,a} \ket{b} \P(\{b\}, \{ a\} )\rho_a\, \bra{b}$.
Thus if we take the state of the instrument to be described by a density matrix,
the diagonal elements of this density matrix, in the basis in which the set
$\{ A_i\}$ are diagonal,  are determined by $\P (\{b\}, \{a\})$.
No information about nondiagonal elements is obtained.
One can consider measurements in different bases, i.e., using different
sets of mutually commuting observables, to identify possible nondiagonal
elements of $\rho$. This is often referred to as
quantum tomography \cite{tomography}.
However, using a different set of compatible observables means that one is using a different set of instruments. If one uses a fixed set of instruments corresponding to a particular choice of the set of compatible observables,
one gets just $\P (\{b\}, \{ a\})$ (for that set).
One may then consider the corresponding (diagonal) $\rho$ as an approximate
description of the state of the instruments after the conclusion of the measurements. One could then use $\rho' = \sum_{b,a} \ket{b} \P(\{b\}, \{ a\} )\rho_a\, \bra{b}$
as the input
density matrix for any subsequent measurements.
The natural question then is:
How good is this $\rho'$ as a description of the state of the instruments
(or the world as seen via the instruments)?

If $\sigma$ denotes the true density matrix of the instrument after
measurements are completed, then a good measure
of how close we are in positing the state to be
described by
$\rho' = \sum_{b,a} \ket{b} \P(\{b\}, \{ a\} )\rho_a\, \bra{b}$ 
is the fidelity given by \cite{fidelity}
\beq
F(\sigma, \rho') = \left[ \Tr \sqrt{\sqrt{\rho'}\, \sigma\, \sqrt{\rho'}}\right]^2
\label{E32a}
\eeq
However, since we do not know the true density matrix, this is
not very useful in the present context.

Another measure is provided by the relative entropy \cite{relentr}.
A series of measurements, say via positive operator-valued measures, to determine the state would naturally lead to
the relative entropy.
(It is the quantum generalization of the Kullback-Liebler distance.)
The determination of the state using measurements of
$\P (\{b\}, \{ a\})$ is a similar strategy.
If $\rho'$
is the diagonal density matrix inferred from the measurements
of $\P (\{b \}, \{ a\} )$ and $\sigma $ is a general nondiagonal
density matrix with eigenvalues $\sigma_k$, i.e., $\sigma = U \sigma_k U^\dagger$ for some unitary matrix $U$, then since 
$\sigma$ gives the same measured values for $\la A_i\ra$, we must have
\beq
\rho'_{b} = \sum_a  \P (\{b\}, \{ a\} ) \rho_a
= \sum_k \vert U_{bk}\vert^2 \sigma_k
\label{E32b}
\eeq
In other words, the diagonal elements of $\sigma = U \sigma_k U^\dagger$
should be the same as $\rho'_b$ to give the same
measured values $\la A_i\ra$ for a complete set of compatible
observables.
The relative entropy of $\sigma$ and $\rho'$ is then given by
\beqar
S_{\rm rel} ( \sigma| \rho') &=& \Tr (\sigma\log \sigma - \sigma \log \rho')
= - \Tr (\rho' \log \rho' ) - \bigl[- \Tr (\sigma \log \sigma )\bigr]\nonumber\\
&=& S(\rho') - S(\sigma)
\label{E32c}
\eeqar
If the trace exists, this is positive. ($\rho'_b$ are linear combinations of
$\sigma_k$ with positive coefficients and since entropy is a concave function,
$S(\rho' ) \geq S(\sigma)$.)
This shows that $\rho'$ which is diagonal has a higher entropy than any
nondiagonal density matrix $\sigma$, which has the same diagonal
elements as $\rho'_b$.
Thus $\rho'$ maximizes the entropy subject to the diagonal values
$\rho'_b$ being determined by $\P(\{b \}, \{ a\})$.
\section{$S$-matrix for a scalar field}
We will now consider simplifying the expression
(\ref{obs5})
for what may be termed as a scalar field theory and
show how it reproduces the usual field theoretic calculations.
We have to make certain approximations which are also inherent in
the usual approaches to the $S$-matrix.

First of all, the initial and final read-outs are
taken to be in the far past and far future, respectively, so that
we may simplify the expression (\ref{obs03}) for $\A$ as
\beq
\A_{\beta b, \alpha a} = \bra{\beta} U^\dagger (t, t_0) f^*_b (t) \,U(t, t_0)\,
f_a (t_0) \ket{\alpha}
\label{obs5a}
\eeq
with $t \rightarrow \infty$, $t_0 \rightarrow -\infty$ eventually.
The probability thus takes the form
\beq
\P(b, a) = \sum_\gamma \sum_\alpha \bra{\gamma} f_b^* U(t, t_0) f_a\ket{\alpha}\, p_\alpha\,  \bra{\alpha} f^*_a U^\dagger (t, t_0) f_b \ket{\gamma}
\label{obs5b}
\eeq
In evaluating this with the functional integral over $A$, $\Phi$,
we get correlations among fields in $U$, separately among
fields in $U^\dagger$ and also cross correlations.\footnote{The result will be similar to the evolution of the density matrix in 
(\ref{E31}), with terms like $T^\alpha_{ac}\rho_{cd} T^\alpha_{db}$ 
representing the cross correlations.}
Usually, one makes the further assumption that the state $\ket{\alpha}$
is restored over the long time-interval so that only one state,
namely $\ket{\gamma} = \ket{\alpha}$ contributes in the sum over
$\gamma$.\footnote{This is not the case for evolution over a finite time-interval, but may be a good approximation for large $(t- t_0)$.}
In calculating the correlations, this state is taken to be the ground (vacuum) state 
for the fields $A$, $\Phi$.
We also take $p_\alpha = 1$ for this state, zero for others, 
so that the starting state and the final state of the world correspond to the
vacuum state of the fields in ambient spacetime. In this case
the sum over $\alpha$ has only one term.
The probability thus reduces to
$\P(b, a)= \vert \A_{ba} \vert^2 $
with
\beq
\A_{ba} = \bra{\alpha} f^*_b (t) U(t, t_0) f_a(t_0) \ket{\alpha}
= \int [DA D\Phi Dx] \, e^{i \S (A, \Phi) } \, I_{ba} (L)
\label{obs5c}
\eeq

To convert this description in terms of instruments 
to the more familiar expression for the $S$-matrix, some further reduction
is needed.
The instrument 
 is taken to be sufficiently massive so that
one can approximate its world line to be
${\dot x}^0 = 1$, ${\dot x}^i = 0$. 
In other words, the functional integrals $[Dx] e^{i \S_{\rm instr}}$
are localized on to these trajectories.
Further, $m^2$ has a dominant term proportional to the identity,
say $m_0^2 \mathbb{1}$,
and the nondiagonal terms corresponding to
the energy differences among the internal states $\ket{a}$
of the instrument (which are equivalent to mass differences
for the instrument as a whole).
Taking the masses to be large, we then approximate
\beq
\sqrt{\Tr[ h U^\dagger (m^2 + \lambda \Phi ) U ]} ~ \sqrt{{\dot x}^2}
\approx  m + {\lambda \Phi \over 2 m_0}
\label{obs6}
\eeq
where in the second term, the mass differences can be neglected.
The nondiagonal terms with energy differences will be kept in the
leading term since they are needed for
the conservation of energy for absorption and emission of
$\Phi$-particles.
Focusing on just the scalar field for now, 
we also neglect the coupling of the gauge field $A_\mu$.
The action for the instrument is thus
\beq
\S_{\rm instr} \approx \int d\tau\, \Tr \left[ h U^\dagger \left(
-i {\del U \over \del \tau} - {\lambda \Phi \over 2 m}U\right) \right]
- m \int d \tau
\label{obs7}
\eeq

A scalar field $\phi$ in the ambient spacetime corresponds to taking
\beq
\Phi  (\vx_p, x_p^0) = (2 m_0) \,\phi(x_p) \, J (x_p)
\label{obs10}
\eeq
where $J (x_p)$ is an operator (or matrix)
which makes the appropriate
transitions $a_p \rightarrow b_p$ for the apparatus.
(The factor of $2m_0$ is just for convenience.)
Some of these transitions can be interpreted as
transitions producing or radiating $\phi$-particles into the ambient space
and some 
correspond to the absorption of the $\phi$-particles.

With these approximations, $\A$ takes the form
\beq
\A_{ba}  = \int [D\Phi D\mu]\, e^{i \S (\Phi )}\, \prod_i f^*_{b_i} e^{i \S_{\rm instr}} f_{a_i}
\label{obs11}
\eeq
We can expand $e^{i \S_{\rm instr}}$ in powers of $\Phi$ and, if we consider
the lowest order term with a nontrivial $a \rightarrow b$ transition
for each of the instruments, (that is we need one factor of
$\Phi$ for each instrument), we get
\beqar
\A_{ba} &=& \int \prod_i dx^0_i \, G(x_M, \cdots, x_1) \, J(x_M)_{b_M a_M}
\cdots J(x_1)_{b_1 a_1} \label{obs12}\\
G(x_M, \cdots, x_1) &=& \int [D\phi] e^{i \S (\phi) } \phi (x_M) \cdots
\phi(x_1) \nonumber\\
J(x)_{ba} &=& \int [D\mu] f_b^* (z) \,e^{i \S_{0,\rm instr}} \, f_a(z')
\label{obs13}\\
\S_{0,\rm instr}&=& \int d\tau\, \Tr \left[ h U^\dagger \left(
-i {\del U \over \del \tau} \right) \right]
- m \int d \tau
\eeqar
The correlation of the $\phi$'s is part of the usual $S$-matrix
element. 
To see this more explicitly, notice that the correlation function
$G(x_M, \cdots, x_1) $ has a factorization of the form
\beq
G(x_M, \cdots, x_1)  = \int_{z_1, \cdots, z_M}
G(x_M, z_M) \cdots G(x_1, z_1) \, V(z_1, \cdots , z_M)
\label{obs13a}
\eeq
where $G(x_i, z_i)$ are propagators for $\phi$, and 
$V(z_1, \cdots , z_M)$ is the vertex function.
By virtue of the choice of the labels
$b$, $a$, the elements $J(x)_{ba} $ pick out the appropriate
terms (referring to the production versus the detection of $\phi$'s) in the mode expansion of the propagators
$G(x_i, z_i)$.
The amplitude $\A_{b a}$ then reduces to the usual
textbook formula.

\section{The Fermi problem}
We will now consider the Fermi problem which has been a much-studied case for signaling and Einstein causality in field theory
\cite{fermi}.
The idea is to consider two atoms, say $A$ and $B$, separated by a
distance $R$. At time $t = 0$, $A$ is in an excited state, $B$ is in its ground state. The question is to calculate the probability $P(t)$
that $B$ is in an excited state at time $t$, to be interpreted as being due to
$B$ absorbing the photon from the spontaneous decay of $A$.
Einstein causality would require that $P(t ) = 0$ for $t < R$ (since it takes time
$R$ for the photon to propagate from $A$ to $B$).
However the simple use of the (time-ordered) Feynman propagator
leads to causality problems since it has correlations over spacelike separations. 
An analyticity argument by Hegerfeldt highlights some of the tricky issues 
involved in the analysis of this seemingly simple problem\cite{heger}, suggesting that either
atom $B$ is never excited or it is in an excited state
even before the decay of atom $A$.
A factorization of the Hilbert space into locally defined Hilbert spaces for
atom $A$ and atom $B$, in the spirit of making local
observations, is the premise of such an argument.
The problem has a long history going back to Fermi's work
in 1932, and has been resolved in different framings from straightforward calculations to the use of algebraic quantum field theory \cite{buchh-yng}.
The factorization of the Hilbert space into locally defined Hilbert spaces for the
atoms is not possible because of the type of von Neumann algebra
for the observables and this can help to resolve the problem.
The direct use of probability, rather than amplitudes, can also be used
to resolve this issue \cite{millington}.
This latter method is not the same but fairly close to
what we show below.
We also refer to \cite{millington} for background historical information.

To set up this problem in our framework, we consider two instruments
corresponding to $A$ and $B$, where we have $I^{(A)}_{ba}$ with the transition $a\rightarrow b$ corresponding to the decay of $A$ and
$I^{(B)}_{dc}$ corresponding to $B$ making a transition
($c \rightarrow d$) to the excited state.
The amplitude (\ref{obs3e}) then takes the form
\beqar
\A_{\beta bd, \alpha ac} &=&
 \int [DA D\Phi Dx D\mu] \braket{\beta| A, \Phi} \,
e^{i\S_{\rm C} (A, \Phi)} \Bigl[f_b^*(w_2^0) \,e^{i \S_{\rm instr}} \, f_a(w_1^0) \nonumber\\
&&\hskip .2in
f_d^*(v_2^0) \,e^{i \S_{\rm instr}} \, f_c(v_1^0)\Bigr]
\braket{A', \Phi'|\alpha}
\label{fermi1}
\eeqar
The actions are over the time-contours as explained in relation to
(\ref{obs3e}). We will consider the possible absorption and emission of
a scalar particle, as this is simpler but captures the essence of the
\begin{figure}[!t]
\begin{center}
\scalebox{1.2}{\includegraphics{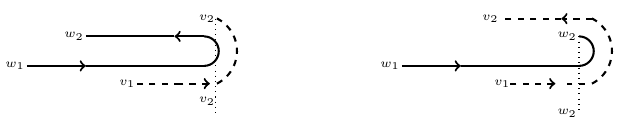}}
\caption{The world lines for the two instruments $A$ and $B$.
The figure on the left applies to $v_2^0 > w_2^0$, the one on the right
to $w_2^0 > v_2^0$. All time-labels for the left figure are $< v_2^0$,
and $< w_2^0$ for the right one; arcs are added only to show continuity of the contours.}
\label{fig2}
\end{center}
\end{figure}
causality issues. This means that we consider a field $\phi$ in the ambient space.
There are two possible world line choices or
time-contours corresponding to
$v_2^0 > w_2^0$ and $w_2^0 < v_2^0$, as shown in Fig.\,\ref{fig2}.
Using the expression for $\A_{\beta bd, \alpha a c}$ and its conjugate
and integrating over $\Phi$ with $\exp( i \S_{\rm SK_4})$, 
we find the probability 
\beq
\P (b d; c a) =  \Bigl\langle
I^{(A)*}_{ba} \,I^{(B)*}_{dc} \, I^{(B)}_{dc} \, I^{(A)}_{ba}\Bigr\rangle
\label{fermi2}
\eeq
The integration over $\phi$, indicated by the angular brackets
in this equation, will produce correlators of $\phi$.
For ease of keeping track of the relevant terms, we can write the
$I$'s in the usual field theory notation 
to the lowest order in $\phi$ as
\beqar
I^{(A)}_{ba} &=& \pm i \int_{w_1^0}^{w_2^0}\bra{b} \rho(w) \ket{a}\, \phi (w)
\nonumber\\
I^{(B)}_{dc} &=& \pm i \int_{v_1^0}^{v_2^0} \bra{d} \rho(v) \ket{c}\, \phi(v)
\label{fermi2a}
\eeqar
where $\rho$ is an operator which effects the transitions
$a \rightarrow b$ and $c\rightarrow d$ in the instruments.
The positive sign in (\ref{fermi2a}) applies when the fields
are on the first branch, the negative sign applies to the second branch.
The instruments are separated by the spatial distance $R$ which is not explicitly
indicated. We can now view the correlations of the fields $\phi$ as given by
Wick contractions. 

In (\ref{fermi2}), we will get four
$\phi$'s to the lowest order, so
we need two correlators for $\phi$
to get a nonzero result in (\ref{fermi2}).
There are then three possibilities given by:
\begin{enumerate}
\item A correlator connecting $I^{(B)}_{dc}$ and $I^{(A)}_{ba}$, with
another correlator connecting
$I^{(A)*}_{ba}$ and $I^{(B)*}_{dc}$, the latter part being the 
conjugate of the first.
\item A correlator connecting $I^{(A)*}_{ba}$  and $I^{(A)}_{ba}$,
with another correlator connecting $I^{(B)*}_{dc}$ and 
$I^{(B)}_{dc}$.
\item A correlator connecting $I^{(B)*}_{dc}$ and $I^{(A)}_{ba}$,
with another correlator connecting $I^{(A)*}_{ba}$ and $I^{(B)}_{dc}$.
\end{enumerate}

We will start by considering the second and third possibilities.
For the second possibility, with the contour ordering we get
$\la \phi ({\tilde v}) \phi(v)\ra$, where $v$ refers to $I^{(B)}_{dc}$
(or $\bra{d} \rho \ket{c}$)
and ${\tilde v}$ to $I^{(B)*}_{dc}$; we will also have a factor
$\la \phi({\tilde w}) \phi (w)\ra$ for $I^{(A)}_{ba}$, $I^{(A)*}_{ba}$.
To be compatible with $B$ making a transition to an excited state
(i.e., for the chosen $d$, $c$ states of the
instrument),
$\phi(v)$ in $I^{(B)}_{dc}$ must correspond to absorption of the $\phi$-particle.
However, $\la \phi({\tilde v}) \phi (v)\ra$ for $\phi(v)$ with the exponential 
$e^{-i \omega t}$ corresponding to absorption is zero, if the ambient spacetime is taken to be in the vacuum state for the field $\phi$.
(If one has a different state for $\phi$, then transitions
like $c\rightarrow d$ are possible even without any input from
atom $A$ and the the formulation of the Fermi problem does not apply.)
Hence the second possibility gives zero.

For the third case, the correlator connecting 
$I^{(B)*}_{dc}$ and $I^{(A)}_{ba}$ will be of the form
$\la \phi ({\tilde v}) \phi (w)\ra$. Both $\phi$'s need to
have a positive exponential of the form $e^{i \omega t}$ to
be compatible with the chosen states, i.e., decay of $A$
and excitation of $B$. The correlator 
$\la \phi ({\tilde v}) \phi (w)\ra$ is zero for this case, so the third
possibility gives zero as well.

Turning to the first possibility, consider the case $v_2^0 > w_2^0$ first.
From Fig.\,\ref{fig2}, we see that $\phi(w)$, corresponding to $I^{(A)}_{ba}$,
can be on the first branch, or the second,
while $\phi(v)$, corresponding to $I^{(B)}_{dc}$,
is on the first branch. For the fields on the first branch, we
get the Feynman propagator $\la \T \phi(w) \phi(v) \ra$
while for $I^{(A)}_{ba}$ on the second branch we get
$\la \phi (w) \phi (v)\ra$.
In the amplitude, we thus get
\beq
\la \T \phi(w) \phi(v) \ra -
\la \phi(w) \phi (v) \ra = \Theta (v^0 - w^0) [\phi(v), \phi(w)]
\label{fermi3}
\eeq
where $\Theta (v^0 - w^0)$ is the step function.
(This is multiplied by the matrix elements for the instruments
as in (\ref{fermi2a}) and a similar 
term for the conjugate.)
The minus sign for the second term arises from the fact that 
$\phi(w) $ in that term is from the time-reversed branch.

In a similar way, for the choice $w_2^0 > v_2^0$, we will get
\beq
\la \T \phi(w) \phi(v) \ra -
\la \phi(v) \phi (w) \ra = \Theta (w^0 - v^0) [\phi(w), \phi(v)]
\label{fermi4}
\eeq
Again, for $\A^*$ we have the conjugate of these expressions.
Since we have the retarded commutators, the result is
compatible with Einstein causality, vanishing over spacelike separations
for the two instruments.
Notice also that we are considering the instruments to be sufficiently
localized in space, and that the time of the read-outs
$w_1^0$, $w_2^0$, etc.
can be unambiguously defined, which means that
the time-separations $w_2^0 - w_1^0$, $v_2^0- v_1^0$
are large compared to the time needed for the transitions
to occur in the instruments.

The square of the retarded commutators is essentially the result 
obtained in
\cite{millington} as well. This paper computed the probability for $B$ to be in any excited state, not a specific one as we have done.
(For us, there was a fixed choice, namely, $d$ for $B$.)
There were, therefore, some additional interference terms
possible in the calculations in \cite{millington} which were crucial
for the result.

\section{Discussion}

There are many situations in physics where
it is necessary to sift out unphysical questions.
For example, in the case of the double-slit interference experiment,
what is measured is a two-instrument
correlation function, namely, the correlation between
the source of the particles and the detector at the screen, both placed
in an ambient spacetime whose geometry carries information about the slits.
Thus, while conjecturally we can consider the question of which slit the particle 
went through, this is not a properly posable question (within the grammar of
quantum mechanics) since there is no detector placed at either slit to
make the question answerable with observations.
(If we place such a detector, it is a different problem of
correlations among three instruments.)
This example again illustrates that it will be useful to have 
a formulation where one is dealing only with 
what is actually observed with the instruments.
This is the main point of this paper.

The dynamics of the instrument then becomes important.
We have given a general form for this dynamics as
a slight generalization of the coadjoint orbit action.
The relevant probabilities are then naturally defined on a four-fold
time-contour. This can be expressed as the correlation function for
a number of Wilson lines for the instruments, so the
framework also ties in with the world-line approach to
calculations in field theory.
As an example to illustrate the method, we considered
the Fermi problem and showed how it is resolved in 
an elementary fashion.

The coadjoint orbit action is a generalization 
of the action used in \cite{witten2} to the case of an
observer/instrument with internal degrees of freedom.
It can thus be applied to situations with a nontrivial background metric for
the ambient spacetime. We plan to investigate this possibility  further
in the near future.

\bigskip
I thank Abhishek Agarwal for some useful comments. 
This work was supported in part by the U.S. National Science Foundation Grant No. PHY-2412479.


\end{document}